# Modeling Spammer Behavior: Naïve Bayes vs. Artificial Neural Networks

Md. Saiful Islam, Shah Mostafa Khaled, Khalid Farhan, Md. Abdur Rahman and *Joy Rahman

Institute of Information Technology, University of Dhaka, Dhaka-1000, Bangladesh
* Services Engineer (IP/MPLS & GPRS Data Network), L M Ericsson Bangladesh Ltd.
E-mail: {saifulit, khaled}@univdhaka.edu, khalidfo@gmail.com, eng_mukul@yahoo.com, joyrahman@gmail.com

*Abstract*—Addressing the problem of spam emails in the Internet, this paper presents a comparative study on Naïve Bayes and Artificial Neural Networks (ANN) based modeling of spammer behavior. Keyword-based spam email filtering techniques fall short to model spammer behavior as the spammer constantly changes tactics to circumvent these filters. The evasive tactics that the spammer uses are themselves patterns that can be modeled to combat spam. It has been observed that both Naïve Bayes and ANN are best suitable for modeling spammer common patterns. Experimental results demonstrate that both of them achieve a promising detection rate of around 92%, which is considerably an improvement of performance compared to the keyword-based contemporary filtering approaches.

*Keywords-component; Spam Email; Machine Learning; Naïve Bayesian Classifier; Artificial Neural Networks*

## I. INTRODUCTION

The exponential growth of junk emails in recent years is a fact of life. Internet subscribers world-wide are unwittingly paying an estimated €10 billion a year in connection costs just to receive spam emails, according to a study undertaken for the European Commission [1]. Though there is no universal definition of spam, unwanted and unsolicited commercial email is basically known as the junk email or spam to the internet community. Spam's direct effects include the consumption of computer and network resources and the cost in human time and attention of dismissing unwanted messages [2]. Combating spam is a difficult job contrast to the spamming. The simplest and most common approaches are to use filters that screen messages based upon the presence of common words or phrases common to junk e-mail. Other simplistic approaches include *blacklisting* (automatic rejection of messages received from the addresses of known spammers) and *whitelisting* (automatic acceptance of message received from known and trusted correspondents). The major flaw in the first two approaches is that it relies upon complacence by the spammers by assuming that they are not likely to change (or forge) their identities or to alter the style and vocabulary of their sales pitches. Whitelisting risks the possibility that the recipient will miss legitimate e-mail from a known or expected correspondent with a heretofore unknown address, such as correspondence from a long-lost friend, or a purchase confirmation pertaining to a transaction with an online retailer. A detail explanation of these techniques is given in [3].

Machine learning algorithms namely Naïve Bayesian classifier, Decision Tree induction, Artificial Neural Networks and Support Vector Machines based on keywords or tokens extracted from the email's *Subject*, *Content-Type* Header and Message *Body* have been used successfully in the past [2-6]. Very soon they fail to filter out spam emails as the spammer changing themselves in the ways that are very difficult to model by simple keywords or tokens [6]. The tricks the spammer uses are themselves patterns that can be modeled to combat spam. Actually the more they try to hide, the easier it is to see them [6]. This study investigates the possibilities of modeling spammer behavioral patterns instead of vocabulary as features for spam email classification. The two machine learning algorithms Naïve Bayes and Artificial Neural Networks are experimented to model common spammer patterns and both of them achieve a promising detection rate that can be considered as an improvement of performance compared to the keyword-based contemporary filtering approaches.

The paper is organized as follows: section 2 discusses the two machine learning algorithms, section 3 presents the features that are modeled, evaluation measures and experimental results and finally section 4 concludes the paper.

## II. MODELING APPROACHES

### A. Naïve Bayesian Classifier

Bayesian classifiers are based on Bayes' theorem. For a training e-mail E, the classifier calculates for each category, the probability that the e-mail should be classified under $C_i$, where $C_i$ is the $i^{th}$ category, making use of the law of the conditional probability:

$$P(C_i|E) = \frac{P(C_i)P(E|C_i)}{P(E)}$$

Assuming class conditional independence, that is, the probability of each word in an e-mail is independent of the word's context and its position in the e-mail, $P(E|C_i)$ can be calculated as the product of each individual word $W_j$'s probabilities appearing in the category $C_i$ ($W_j$ being the $j^{th}$ of $l$ words in the e-mail):

The category maximizing $P(E|C_i)$ is predicted by the classifier [5] [7].

## B. Artificial Neural Networks

Artificial neural networks (ANN) are non-linear statistical data modeling tools that tries to simulate the functions of biological neural networks. It consists of interconnected collection of simple processing elements or artificial neurons and processes information in a connectionist approach to computation [6] [7]. ANN is generally considered to be an adaptive system that changes its structure in response to external or internal information that flows through the network during the learning phase. Fig 1 shows an example of multilayer feed forward neural network.

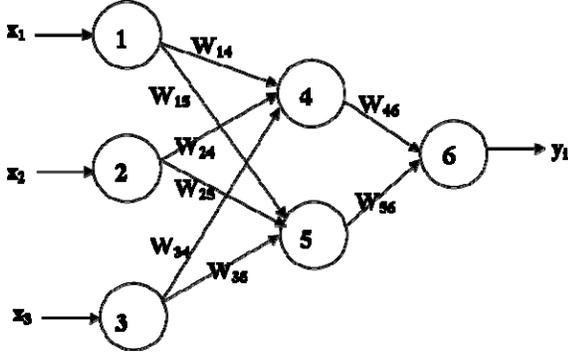

Figure 1. An example of a multilayer feed-forward artificial neural network.

Neural Networks have long training time and require a large number of parameters that are best determined empirically. Neural Networks have been criticized for their poor interpretability since it is difficult to interpret the symbolic meaning behind the learned weights [7].

## III. EXPERIMENTAL SETUP AND RESULTS

In this work a corpus of 200 emails received by one of the authors over a period of several months is used for experimentation. The distribution of both spam and non-spam emails in this collection is equal. That is, out of 200 emails 100 is spam and 100 is non-spam.

Each email is parsed as text file to identify each header element to distinguish them from the body of the message. Every substring within the subject header and the message body that was delimited by white space was considered to be a *token*, and an *alphabetic word* was defined as a token delimited by white space that contains only English alphabetic characters (A-Z, a-z)or apostrophes. The tokens were evaluated to create a set of 18 hand-crafted features from each e-mail message (Table 1) of which features 1-17 are proposed in [6]. In addition of these 17 features this study proposes other four features 18-21. The study investigates the suitability of these 21 features in classifying spam emails.

Estimating classifier accuracy is important since it allows one to evaluate how accurately a given classifier will classify unknown samples on which the classifier has not been trained. The effectiveness of a classifier is usually measured in terms of accuracy, precision and recall [5] [7]. These measures are calculated using the confusion matrix given below:

| Category $C_i$ | Correct | |
|---|---|---|
| Predicted ↓ | YES | NO |
| YES | $TP_i$ | $FP_i$ |
| NO | $FN_i$ | $TN_i$ |

TP = true positives
FP = false positives
FN = false negatives
TN = true negatives

TABLE I. FEATURES EXTRACTED FROM EACH EMAIL

| Feature | Category 1: Features From the Message Subject Header |
|---|---|
| 1 | Binary feature indicating 3 or more repeated characters |
| 2 | Number of words with all letters in uppercase |
| 3 | Number of words with at least 15 characters |
| 4 | Number of words with at least two of letters J, K, Q, X, Z |
| 5 | Number of words with no vowels |
| 6 | Number of words with non-English characters, special characters such as punctuation, or digits at beginning or middle of word |
| | **Category 2: Features From the Priority and Content-Type Headers** |
| 7 | Binary feature indicating whether the priority had been set to any level besides normal or medium |
| 8 | Binary feature indicating whether a content-type header appeared within the message headers or whether the content type had been set to "text/html" |
| | **Category 3: Features From the Message Body** |
| 9 | Proportion of alphabetic words with no vowels and at least 7 characters |
| 10 | Proportion of alphabetic words with at least two of letters J, K, Q, X, Z |
| 11 | Proportion of alphabetic words at least 15 characters long |
| 12 | Binary feature indicating whether the strings "From:" and "To:" were both present |
| 13 | Number of HTML opening comment tags |
| 14 | Number of hyperlinks ("href=") |
| 15 | Number of clickable images represented in HTML |
| 16 | Binary feature indicating whether a text color was set to white |
| 17 | Number of URLs in hyperlinks with digits or "&", "%", or "@" |
| 18 | *Number of color element (both CSS and HTML format)* |
| 19 | *Binary feature indicating whether JavaScript has been used or not* |
| 20 | *Binary feature indicating whether CSS has been used or not* |
| 21 | *Binary feature indicating opening tag of table* |

Accuracy of a classifier is calculated by dividing the number of correctly classified samples by the total number of test samples and is defined as:

$$Accuracy = \frac{number\ of\ correctly\ classified\ samples}{total\ number\ of\ test\ samples}$$

$$= \frac{TP+TN}{TP+FP+FN+TN}$$

Precision measures the system's ability to present only relevant items while recall measures system's ability to present all relevant items. These two measures are widely used in TREC evaluation of document retrieval [9]. Precision is calculated by dividing the number of samples that are true positives by the total number of samples classified as positives and is defined as:

$$\Pr ecision = \frac{number\ of\ true\ positives}{total\ number\ of\ samples\ classified\ as\ positives}$$

$$= \frac{TP}{TP+FP}$$

Analogously, recall is calculated by dividing the number of samples that are true positives by the total number of samples that classifier should classified as positives and is defined as:

$$\text{Re}call = \frac{number\ of\ true\ positives}{total\ number\ of\ positive\ samples}$$

$$= \frac{TP}{TP+FN}$$

In this study, both precision and recall are kept close to give equal importance on both of them. Table 2 summarizes the comparative results of the two well-known machine learning algorithms namely Naïve Bayesian classifier and Artificial Neural Networks. These algorithms are tested on Weka 3.6.0 suite of machine learning software written in Java, developed at the University of Waikato [10]. It is observed that Naïve Bayesian classifier outperforms than ANN learning algorithms in all cases. The highest level of accuracy that can be achieved by Naïve Bayesian classifier is 92.2% (shown in yellow color in Table 2) using features from category 2 and 3. The accuracy that can be achieved by any learning algorithms using features from category 1 is negligible. Features from category 2 and 3 contribute mostly in classifying spam emails from non-spam emails for all machine learning algorithm experimented in this study.

Highest number of features is always desirable only if their inclusion increase classifier's accuracy significantly. Growing number of features not only hinders multidimensional indexing but also increases overall execution time. So, this study starves to find an optimal number of features that can be effectively used to lean a classifier without degrading the level of accuracy.

Applying best first forward attribute selection method the study gets only 10 features from category 2 and category 3 useful for classifying the spam and non-spam emails without sacrificing the accuracy as shown in Table 3. The set includes features 8, 9, 10, 12, 13, 14, 15, 16, 17, and 18 of which feature 18 is identified in this study. The Naïve Bayesian classifier again outperforms than ANN learning algorithms. The optimal feature set obtained by applying best first forward attribute selection method for the features proposed in [6] includes only features 8, 9, 10, 12, 13, 14, 15, 16 and 17, a total of 9 features. In this case ANN outperforms than Naïve Bayesian classification algorithm (shown in light blue in Table 3).

IV. CONCLUSION

This paper studies the modeling of spammer behavior by the two well-known machine learning algorithms for spam email classification. Based on examining different features and two different learning algorithms, the following conclusions can be drawn from the study presented in this paper:

- Spammer behavior can be modeled using features extracted from Content-Type header and message Body only.
- The contribution of features extracted from subject header in spam email detection is negligible or insignificant.
- Naïve Bayesian classifier models the spammer behavior best than Artificial Neural Networks.

It is possible to get an optimal number of features that can be effectively applied to learning algorithms to classify spam emails without sacrificing accuracy.

REFERENCES

[1] "Data protection: "Junk" e-mail costs internet users 10 billion a year worldwide - Commission study", Last Accessed on 14[th] Feb, 2009. http://europa.eu/rapid/pressReleasesAction.do?reference=IP/01/154

[2] M. Aery and S. Chakravarthy, "eMailSift: email classification based on structure and content", In Proc. of 5[th] IEEE Intl. Conf. on Data Mining, 2005, pp. 1-8.

[3] Md. Rafiqul Islam and M. U. Chowdhury, "Spam filtering using ML algorithms", In Proc. of IADIS International Conf. on WWW/Internet, 2005, pp. 419-426.

[4] H. Drucker, D. Wu, and V. N. Vapnik, "Support vector machines for spam categorization", IEEE Transactions on Neural Networks, Vol. 10, No. 5, 1999, pp. 1048-1054.

[5] K. Eichler, "Automatic Classification of Swedish Email Messages", B.A Thesis, Eberhard-Karls-Universitat Tubingen, 17[th] August, 2005.

[6] I. Stuart, S. Cha, and C. Tappert, "A neural network classifier for junk E-mail", Lecture notes in computer science, Vol. 3163, 2004, pp. 442-450.

[7] J. Han and M. Kamber, Data Mining Concepts and Techniques, Academic Press, ISBN 81-7867-023-2, 2001.

[8] Md. Saiful Islam and Md. Iftekharul Amin, "An architecture of active learning SVMs with relevance feedback for classifying E-mail", Journal of Computer Science, Vol. 1, No. 1, 2007, pp. 15-18.

[9] J. Makhoul, F. Kubala, R. Schwartz and R. Weischedel, "Performance measures for information extraction", In Proc. of DARPA Broadcast News Workshop, Herndon, VA, February 1999.

[10] G. Holmes; A. Donkin and I.H. Witten, "Weka: A machine learning workbench", In Proc. 2nd Australia and New Zealand Conference on Intelligent Information Systems, Brisbane, Australia, 1995.

[11] S. Youn and D. McLeod, "A Comparative Study for Email Classification", Advances and Innovations in Systems, Computing Sciences and Software Engineering, pp. 387-391, 2007.

TABLE II. COMPARISON RESULTS FOR NAÏVE BAYESIAN CLASSIFIER AND ARTIFICIAL NEURAL NETWORK

| Features | Naïve Bayesian Classifier(NaiveBayes) | | | ANN (Multilayer Perceptron) | | |
|---|---|---|---|---|---|---|
| | Accuracy | Precision | Recall | Accuracy | Precision | Recall |
| Category 1 Only | 56.5 % | 55.7% | 56.5% | 67.8% | 73.1% | 67.8% |
| Category 2 Only | 65.2% | 75.0% | 65.2% | 65.2% | 75.0% | 65.2% |
| Category 3 Only | 88.7 % | 88.7% | 88.7% | 86.1% | 86.1% | 86.1% |
| Category 1+Category 2 | 66.9 % | 67.3% | 67.0% | 73.1% | 77.2% | 73.0% |
| Category 2+ Category3 | 92.2 % | 92.2% | 92.2% | 87.8% | 88.1% | 87.8% |
| Category 1+Category 3 | 80.8 % | 80.9% | 80.9% | 74.7% | 75.4% | 74.8% |
| Category1+ Category 2 + Category 3 | 86.9 % | 87.0% | 87.0% | 84.3% | 85.1% | 84.3% |

TABLE III. EVALUATION OF NAÏVE BAYES AND ANN WITH OPTIMAL FEATURE SET

| Features | Naïve Bayesian Classifier(NaiveBayes) | | | ANN (Multilayer Perceptron) | | |
|---|---|---|---|---|---|---|
| | Accuracy | Precision | Recall | Accuracy | Precision | Recall |
| Best first: 8, 9, 10, 12, 13, 14, 15, 16, 17, and 18 [This study] | 92.2% | 92.2% | 92.2% | 90.4% | 90.6% | 90.4% |
| Best first: 8, 9, 10, 12, 13, 14, 15, 16, and 17 [6] | 86.1 % | 87.4% | 86.1% | 91.3% | 91.4% | 91.3% |